\documentclass[bookmarksopen,twoside]{article}
\usepackage{graphicx,amssymb,mathrsfs,amsmath,color,fancyhdr}
\usepackage[bookmarks=true,dvipdfm]{hyperref}
\input vatola.sty \input cyracc.def
\input psfig.sty

\TagsOnRight

\renewcommand{\baselinestretch}{1.06} 
\catcode`@=11 \long\def\@makefntext#1{\noindent #1}
\newskip\tabcentering \tabcentering=1000pt plus 1000pt minus 1000pt
\def\REF#1{\par\hangindent\parindent\indent\llap{#1\enspace}}
\def\MCH#1#2{\setbox0=\hbox{\raise#1\hbox{#2}}\smash{\box0}}

\def\dl{\displaystyle}
\let\@oddfoot\@empty  \let\@evenfoot\@empty

\def\@evenhead{}\def\@oddhead{}
\def\@evenhead{\vbox{\hbox to \textwidth{\footnotesize\rm\hbox to
1.0cm{\thepage\hfill} \hfill\hspace{2mm}\footnotesize{
\emph{ZHANG Q. F. and CUI J. Z.}}}}}
\def\@oddhead{\vbox{\hbox to \textwidth{\footnotesize
{\it Global well-posedness for Rosseland equation} \hfill{\ } \hfill\hbox to
1cm{\hfill\thepage}}}}

\def\sec#1{\vspace{2mm}\noindent{{\bf #1}}\vspace{0.5mm}}
\def\th#1{\vspace{1mm}\noindent{\bf #1}\quad } 

\def\leq{\leqslant}
\def\geq{\geqslant}
       
  \def\hml{\end{document}}  \newsymbol\wjzhml 203F \def\no{\noindent}

\begin{document}
\abovedisplayskip=3pt plus 1pt minus 1pt 
\belowdisplayskip=3pt plus 1pt minus 1pt 

\def\le{\leqslant}
\def\ge{\geqslant}
\def\dl{\displaystyle}




\vspace{8true mm}

\renewcommand{\baselinestretch}{1.9}\baselineskip 19pt
\begin{center}
\noindent{\LARGE\bf Existence of Rosseland equation}

\vspace{0.5 true cm}

\noindent{\normalsize\sf ZHANG QiaoFu$^{\dag}$ \& CUI JunZhi
\footnotetext{\baselineskip 10pt
$^\dag$ Corresponding author\\
}}

\vspace{0.2 true cm}
\renewcommand{\baselinestretch}{1.5}\baselineskip 12pt
\noindent{\footnotesize\rm $
 Academy\, of\, Mathematics\, and\, Systems\, Science, \,
Chinese\, Academy\, of \,Sciences, \,Beijing\, 100190,\,China
$ \\
(email: zhangqf@lsec.cc.ac.cn,\,cjz@lsec.cc.ac.cn)\vspace{4mm}}
\end{center}

\baselineskip 12pt \renewcommand{\baselinestretch}{1.18}
\noindent{{\bf Abstract}\small\hspace{2.8mm} 
The global boundness, existence and uniqueness are presented for the kind of Rosseland equation with a small parameter. This problem  comes from  conduction-radiation coupled heat transfer in the composites; it's with coef\mbox{}f\mbox{}icients of high order growth and  mixed boundary conditions.
 A linearized map is constructed by f\mbox{}ixing the function variables in the coef\mbox{}f\mbox{}icients and the right-hand side.
The solution to the linearized problem is uniformly bounded based on De Giorgi iteration; it is bounded in the H\"older space from a Sobolev-Campanato estimate.
 This linearized map is compact and continuous  so that there exists a  f\mbox{}ixed point.
  All of these estimates are independent of the small parameter.
 At the end, the uniqueness of the solution holds if there is a big zero-order term and the solution's gradient is bounded.
 This existence theorem can be extended to the nonlinear parabolic problem.
}

\vspace{1mm} \no{\footnotesize{\bf Keywords:\hspace{2mm}}
nonlinear elliptic equation, well-posedness, f\mbox{}ixed point, mixed boundary conditions, without growth conditions, Rosseland equation
}

\no{\footnotesize{\bf MSC(2000):\hspace{2mm} } 35J60,\,47H10
 \vspace{2mm}
\baselineskip 15pt
\renewcommand{\baselinestretch}{1.22}
\parindent=10.8pt  
\rm\normalsize\rm

\sec{1\quad Introduction}
\newcommand{\dif}{\,\mathrm{d}}
\setcounter{section}{1}
\renewcommand\theequation{\arabic{section}.\arabic{equation}}

\newcounter{bh}
\usecounter{bh}
\newenvironment{remark}{\arabic{section}.\refstepcounter{bh}\arabic{bh}.\ }{}
\vspace{2mm}
\noindent Our original motivation is the Rosseland equation in the conduction-radiation  coupled  heat transfer [1,2].
F\mbox{}ind  $(u_\varepsilon-u_b) \in W^{1,2}_0(G)$ (Def\mbox{}inition 2.3), such that
 \begin{equation*}
\int_G a_{ij}(u_\varepsilon(x),x,\frac x \varepsilon)\frac{\partial u_\varepsilon}{\partial x_i}\frac{\partial \varphi}{\partial x_j}+\int_\Gamma \alpha(u_\varepsilon-u_{gas})\varphi
=\int_G f(u_\varepsilon(x),x,\frac x \varepsilon)\varphi,\quad \forall \varphi\in W^{1,2}_0(G),
\end{equation*}
 where $a_{ij}=k_{ij}(x,\frac x \varepsilon)+4u_\varepsilon^3b_{ij}(x,\frac x \varepsilon)$;  $(k_{ij}),(b_{ij})$ are symmetric positive def\mbox{}inite; $k_{ij}(x,y),b_{ij}(x,y)$ are 1-periodic in $y$. The small
parameter $\varepsilon$ is the period of the composite structure. $\Gamma$ is the natural boundary part of $\partial G$.
There may be no ellipticity for $A=(a_{ij})$ without considering physical conditions;  uniform estimates independent of $\varepsilon$
 are also needed. This open problem (the existence theory for the equation with coef\mbox{}f\mbox{}icients like $k+4u^3b$, without $\varepsilon$)
was proposed by Laitinen  in 2002 (Remark 3.4 [3]).

There are several steps: f\mbox{}irstly describe the  physical conditions and f\mbox{}ind a suitable temperature interval by the global boundness in $L^\infty$ (Lemma 3.1); then construct a  linearized map with a f\mbox{}ixed point in this interval (Theorem 3.4);  the f\mbox{}ixed point is unique if there is a big zero-order term and the solution's gradient is bounded (Theorem 4.3).

 The novelty is we don't need any growth conditions in [4]: this method can be used for  coef\mbox{}f\mbox{}icients like $k+4u^mb$, $\forall m>0$. More specif\mbox{}ically,
 $\forall \,C_1, C_2\,\,, 0< C_1\leq C_2$,
 \begin{equation}\label{}
A(u_\varepsilon(x),x,\frac x \varepsilon)\in  [C_3,C_4],\quad\textrm{if}\,\, u_\varepsilon\in [C_1,C_2];\quad 0<C_3=C_3(C_1),C_4=C_4(C_1,C_2).
\end{equation}

 Our main tool is the regularity established by
Griepentrog and  Recke in the Sobolev-Campanato space [5].  Their work asserted that linear elliptic equation of second order with non-smooth
data ($L^\infty$-coef\mbox{}f\mbox{}icients, Lipschitz domain, regular sets, non-homogeneous mixed boundary
conditions) has a unique solution in $C^\beta(\overline{\Omega})$; this H\"older norm smoothly depends  on the data.

Note that the well-posedness is still valid  if the ellipticity is \textit{a priori} known
 or only Dirichlet boundary condition is  considered (the famous De Giorgi-Nash estimate holds; see Theorem 8.29 [4]).
 We present a local gradient estimate for a simplif\mbox{}ied problem (only the righthand side is nonlinear) in Lemma 4.1; it can be  used in the error estimate of the homogenization [6].
 All of these results can be extended to the nonlinear parabolic equation if we use  the regularity in the parabolic Sobolev-Morrey space [7].

Throughout this paper, $C,C_i$ denote  positive constants independent of the solution and the small parameter $\varepsilon$.
The unit cell $Y=(0,1)^n$. $B(x,r)$ is the open ball of radius $r$ centered at $x$.
$\varphi\in [C_0,C_1]$ means that
  $\varphi \in L^\infty$ in the relevant domain if without confusion and $C_0\leq \varphi\leq C_1$.
  For a real symmetric matrix $A=(a_{ij}(u(x),x,y))$,   $A\in [C_0,C_1]$ implies
\begin{equation*}
  a_{ij}(u(x),x,y)\xi_i\xi_j\geq C_0|\xi|^2,\quad\sum|a_{ij}(u(x),x,y)|^2\leq C_1^2
  ,\quad\forall (x,y)\in \Omega\times Y\,,\xi\in \mathbb{R}^n.
\end{equation*}

  $\|\varphi\|_q$ is an abbreviation of the norm in the relevant
$L^q$ space. $T_{min},T_{max}$ are positive physical constants (the range of the environmental temperature); $0<T_{min}\leq T_{max}$.

\vspace{5mm}

\sec{2\quad Regular sets, Campanato space and model problem}
\setcounter{equation}{0}
\setcounter{section}{2}
\setcounter{bh}{0}
\vspace{2mm}

\vspace{5mm}

 \sec{5\quad Conclusions}

\vspace{2mm}
The well-posedness is given for the Rosseland equation with a small parameter $\varepsilon $.
The physical conditions   are included in (A1)-(A5).
Based on the boundness in $L^\infty$, we construct a closed convex set $[T_{min}, T_*]$.
 Then, we prove the linearized map is compact and continuous from the Sobolev-Campanato estimate established by  Griepentrog and Recke.
 So there exists a f\mbox{}ixed point; the solution to the original nonlinear problems  has almost the same estimates as the linear one.
  These estimates are independent of the small parameter. So there is a subsequence which converges in $C^0(\overline{\Omega})$ (or $H^1(\Omega)$), if $\varepsilon\rightarrow 0$.
    A local gradient estimate of the solution is given for a simplif\mbox{}ied problem; it can be used to the error estimate of the same type of equation's  homogenization.
 The uniqueness is also based on a linearized map; see \eqref{eq:linea}.  Similar results on the nonlinear parabolic problem based on the same method and Sobolev-Morrey estimate [7] will appear elsewhere.

\vspace{3mm}
\th{Acknowledgements}
This work is supported by National Natural Science Foundation of China (Grant No. 90916027).
The authors thank   the referees for their careful reading and helpful comments.

\vskip0.1in \no {\normalsize \bf References}
\vskip0.1in\parskip=0mm \baselineskip 15pt
\renewcommand{\baselinestretch}{1.15}

\footnotesize\parindent=6mm

 \REF{1\ }Zhang Q. F., Cui J. Z.,
Multi-scale analysis method for combined conduction-radiation heat transfer of periodic composites,
 Advances in Heterogeneous Material Mechanics(eds. Fan J. H., Zhang J. Q., Chen H. B., \textit{et al}), Lancaster: DEStech Publications, 2011, 461-464

 \footnotesize\parindent=6mm
 \REF{2\ }Modest M. F.,  Radiative heat
transfer, 2nd,  San Diego: McGraw-Hill, 2003

\footnotesize\parindent=6mm
 \REF{3\ }Laitinen M. T.,  Asymptotic analysis of
conductive-radiative heat transfer,   Asymptotic Analysis, 2002,
29(3): 323-342

\footnotesize\parindent=6mm
 \REF{4\ }Gilbarg D., Trudinger N. S., Elliptic Partial Differential Equations  of Second Order, Berlin: Springer, 2001

 \footnotesize\parindent=6mm
 \REF{5\ }Griepentrog J.A., Recke L., Linear elliptic boundary value problems with non-smooth
data: normal solvability on Sobolev-Campanato spaces,  Math Nachr, 2001, 225(1): 39-74

\footnotesize\parindent=6mm
 \REF{6\ }Zhang  Q. F., Cui  J. Z., Regularity of the correctors and local
gradient estimate of the  homogenization for the elliptic equation: linear periodic case, 2011, arXiv:1109.1107v1 [math.AP]

\footnotesize\parindent=6mm
 \REF{7\ }Griepentrog  J.A.,  Sobolev-Morrey spaces associated with evolution equations,
   Adv Differential  Equations, 2007,  12: 781-840

\footnotesize\parindent=6mm
 \REF{8\ }Gr\"oger  K,  A $W^{1,p}$-estimate for solutions to mixed boundary value problems for second order elliptic differential equations, Math Ann, 1989, 283(4): 679-687

\footnotesize\parindent=6mm
 \REF{9\ }Brezis  H., V\'azquez  J. L.,
 Blow-up solutions of some nonlinear elliptic problems, Rev Mat Univ Complut Madrid, 1997, 10(2): 443-469

 \footnotesize\parindent=6mm
 \REF{10\ }Wu  Z. Q., Yin  J. X. and Wang  C. P., Elliptic and Parabolic Equations, Singapore: World Scientif\mbox{}ic, 2006

\footnotesize\parindent=6mm
 \REF{11\ }Bensoussan A., Lions J. L., Papanicolaou G.,
Asymptotic Analysis for Periodic Structures, Amsterdam: North-Holland, 1978

\footnotesize\parindent=6mm
 \REF{12\ }Avellaneda M., Lin F. H., Compactness method in the theory of homogenization,
Comm Pure Appl Math, 1987, 40(6): 803-847

\footnotesize\parindent=6mm
 \REF{13\ }Kenig  C. E.,  Lin  F. H. and  Shen  Z. W.,
 Homogenization of elliptic systems with Neumann boundary conditions, 2010,
 arXiv: 1010.6114v1 [math.AP]

 \hml